\documentclass[preprint,
 amsmath,amssymb,
 aps,
]{revtex4-1}
 \usepackage[colorlinks=true, allcolors=blue]{hyperref}
\usepackage{graphicx}
\usepackage{dcolumn}
\usepackage{bm}
\usepackage[draft,commentmarkup=uwave]{changes} 
\definechangesauthor[color=green!50!black]{Alan}
\usepackage{lipsum}
\usepackage{subcaption}
\usepackage{xcolor}
\usepackage{hhline,booktabs}
\newcommand{\sh} {/ \hskip-5pt }
\begin{document}
 
\title{Pion Valence Structure at Intermediate x in the Residual Field Approach}
\author{Joseph Maerovitz,   Misak Sargsian,}
\affiliation{Florida International University, Miami, FL 33199, USA}
\author{Christopher Leon }
\affiliation{Los Alamos National Laboratory, Los Alamos, NM 87545, USA }
 
\date{\today} 

\begin{abstract}
We calculate the valence parton distribution function~(PDF) of  pion, 
 within the theoretical approach based on spectral function representation of valence quarks in 
the pion. In this approach we assume that the soft partonic structure of pion is defined by 
the valence $q\bar q$ cluster, consisting of current quarks, embedded in the residual field of the pion.
The valence PDF  is calculated using  phenomenological Light-Front  wave functions for the  $q\bar q$ 
cluster and the residual field.  Our result indicates that the peak position of the  x weighted valence PDF~(xPDF)  
depends on parameters characterizing the virtuality of the cluster and the mass of the residual system. 
Magnitudes of these parameters are obtained  by fitting to the height and the peak position of  empirical pion xPDF evaluated at starting $Q_0$. They indicate that unlike the nucleon case, very little  residual mass is needed to describe existing pion PDFs at intermediate x. They also indicate that the $q\bar q$-cluster is highly virtual and on average the interacting quark carries almost all of the momentum of the $q\bar q$ cluster. This picture is consistent with  the dominance of the Feynman mechanism leaving little room for the hard component in the PDF, and practically describing $\sim (1-x)$ behavior observed recently at $x\to 1$ limit. Our non-trivial observation is that the  height and the peak position of xPDF define the analytic behavior of valence PDF at $x\to 1$ limit.

\end{abstract}

\maketitle

\section{Introduction}

As the lightest hadron and a pseudo-Goldstone boson of chiral symmetry breaking, the pion holds a unique position in hadronic and nuclear physics. Studying its structure is essential for understanding the nature of the
strong interaction of nucleons  which in part is mediated through the  pion exchange. 

While in the chiral symmetry breaking picture the long range part of the nucleon-nucleon~(NN) interaction is 
mediated through the near-massless Goldstone field without apparent  internal structure of pion, the deep-inelastic scattering from pion indicates its  underlying quark-gluon structure which does not reveal itself in NN interaction. 
This apparent duality indicates on possible specific distribution 
of valence quarks in pion that makes for example the quark-interchange in NN 
interaction through the  pion field highly unlikely.

One of the important quantities that allows to explore the 
dynamical structure of pion is its parton distribution function (PDF) probed 
at the light-front momentum fraction, x ($0<x<1$) and given resolution $Q^2$. 
It allows us to probe valence and sea quark as well as gluon 
distributions which dominate at different regions of $x$ in the pion.  
For the valence quarks pion represents the 
simplest system and as such it is the best testing ground for 
understanding the QCD mechanism of valence quark interaction in the hadron,
thus attracting  multitude of theoretical studies 
(see e.g. \cite{Brodsky:1985qs,Frederico:1994dx,Tiburzi:2002tq,Brodsky:2007hb,Holt:2010vj,Accardi:2009br,Pasquini:2023aaf,Ding:2019lwe,Gao:2020ito})
.

Much progress has been made in the extraction of pion parton distribution  exploring 
variety of deep-inelastic reactions including Drell-Yan, $J/\Psi$- production and Sullivan processes (see. e.g. \cite{Owens:1984zj,Gluck:1991ey,
Sutton:1991ay,Gluck:1999xe,Wijesooriya:2005ir,Aicher:2010cb,ZEUS:2002gig,H1:2010hym,Barry:2018ort, Barry:2021osv,Barry:2018ort, Barry:2021osv}).

The recent high quality combined analysis of Drell-Yan and Sullivan processes\cite{Barry:2018ort, Barry:2021osv} produced an unexpected result  for  PDFs at  large  $x$, observing a behavior of  $(1-x)^\beta$, with $\beta\approx 1-1.2$.  
This contradicts to the generally expected dependence of $(1-x)^2$ which in part follows from the expectation that high x parton distributions are generated through the hard gluon exchange between valence quarks\cite{farrar1975pion, berger1979quark, Gunion:1983ay}.

Focusing on the dynamics of valence quark distribution in hadrons, 
we notice two important properties of valence quarks that will be used in 
our calculation. The first, is that the valence quarks define quantum numbers of 
hadrons and their flavor is conserved. As such they can be considered as effective fermions whose number is conserved. The second prominent feature of valence quarks 
is that their distribution weighted by the momentum fraction $x$, $xf_i(x)$
shows a distinctive peak associated with the average momentum fraction 
carried by valence quarks.  In general, such a peaking property in  Fermi systems is associated with the ``bulk" property of  bound system which has an inherently non-perturbative character.
In our recent works\cite{Leon:2020cev,Leon:2020bvt} on modeling valence quark  distributions in the nucleon 
the ``bulk" property of hadrons have been described  
through the quark-gluon system residual to the valence quarks in the nucleon. 
The model is  based on  residual mean field approximation which introduces the Light-Front wave function of the
valence quark cluster embedded in the interacting residual field.
Within this model for the case of the nucleon it was found that at the 
starting $Q^2_0= m_c^2$ ($m_c$ is the mass of the charm quark) when one expects the partonic  picture to be relevant, 
the $xf_i(x)$ for valence quark PDFs of nucleon peaks at 
$x_p\le {1\over 4}$ not at ${1\over 3}$ as naively expected. This observation is 
confirmed by the modern phenomenological PDFs of nucleon. 
By describing  these  peaking properties of the valence PDFs of nucleons, we found that it requires the residual  system to have a mass comparable to the pion mass. 
As such the model indirectly favored the existence of the structure in 
the nucleon resembling a pion cloud. The model descried reasonably well the valence PDFs for the region $0.2 < x <0.6$ underestimating  at larger x which indicates  
the importance of hard mechanism in generation of large x component of valence PDFs in the nucleon.\\
\indent In the current work we extend the residual field approach for the description of 
pion valence PDFs. Our goal is to address following main questions: are there 
structure in pion residual to $q\bar q$ valence quarks and  how well  
the mean field distribution of the residual system describes the high x 
structure of pion PDFs? We introduce new object, that is Light-Front wave function of 
the residual system which can be investigated further in probing pion form-factors, 
generatlized partonic and transverse momentum distributions as well as in processes  
involving  target fragmentation in collider kinematics~\cite{Abir:2023fpo}.

The outline for this work is as follows: Sec. \ref{theoretical-framework} provides the description of 
the theoretical framework. In Sec. \ref{pion-in-RMFM} we derive the PDFs 
based on the  residual field model and in Sec.\ref{Sec:Model_LFWF} we introduce light-front wave function of two-body system, 
applying it for  $q\bar q$-cluster and valence-residual systems respectively.
Sec. \ref{fitting-results}  presents numerical estimates  of the model parameters following from 
the fit of theoretical expression to the recently obtained  
phenomenological parameterization of pion PDFs in  the range of $0.1\le x\le 0.85$. Our focus in 
this comparison is to evaluate the relevant masses  and the light-front momentum distribution of residual system 
in the pion through the description of peaking property of x weighted pion PDfs at starting $Q_0$.
We this section we  present also the interpretation of the results, Sec. \ref{conclusion-and-outlook} contains 
conclusions and outlook for further application of the presented approach in different processes.

\section{Theoretical Framework of Residual Field Model}\label{theoretical-framework}

The residual field model utilizes conservation of valence parton quantum numbers to treat them as effective fermions whose numbers are conserved. It separates hadrons 
into a core of valence fermions consisting of massless quarks and a residual structure which contains everything besides the valence quarks, such as sea quarks and gluons. 
Interaction with the external probe in this case proceeds through the 
scattering off the  valence quarks in the cluster.
This picture is assumed to be valid for moderate to high x, $x \gtrsim 0.1-0.2$ and will break down  
at lower $x$, where the Regge dynamics dominate  in the interaction.

According to the residual field model the deep inelastic scattering from a hadron in the leading approximation can be described by the Light-Front diagram of 
Fig.(\ref{fig:rfm_diagrams})(a) in which scattering takes place with a quark inside the valence cluster. The consideration  of the residual system in the scattering amplitude is equivalent  to the spectral function representation of the partonic structure of the hadron in which case the valence quark distribution is indirectly related to the dynamical characteristics of  the residual system.

In light front (LF) quantization one probes quantum fields at fixed light-cone time, $x^+ = t +z$, which provides several simplifications in describing high energy processes \cite{brodsky1998quantum}. The LF approach provides a framework to study hadrons in terms of their light front wave functions (LFWFs), in 
which case a hadron in a ket state represented by $|h\rangle$ can be described through the expansion in terms of the LFWFs ($ \psi_{i/h}\left(\{x_i, \mathbf{k}_i\}_i\right) $) \cite{brodsky1998quantum, kovchegov2012quantum}:
\begin{equation}\label{Fock-expansion}
| h \rangle = \sum_i \int [d\mu_i]  \psi_{i/N}\left(\{x_j, \mathbf{k}_j\}_j\right) |i \rangle,     
\end{equation}
where the sum is over all Fock states $|i\rangle$ and $[d\mu_i] $ is the relevant integration measure. Color and 
spin degrees of freedom have been suppressed. 
 
With above wave functions the PDF for a quark of flavor $i$,  in hadron, $h$  can be represented as the 
expectaion value of the  number operator in the hadronic state \cite{sterman1995handbook,Belitsky:2005qn}:
\begin{equation}
f_i(x,Q^2) = \int {d^2 {\bf k_\perp} \over (2\pi)^2}\langle h(p)\mid b_i^\dagger(xp,{\bf k_\perp})  b_i(xp,{\bf k_\perp})\mid h(p)\rangle,
\label{eq.pdfDefinition2}
\end{equation}
where $b^\dagger_i$ and $b_i$ are creation and annhiliation operators of parton $i$.

Except for the situations of high $Q^2$ and  $x\to 1$ in which one expects minimal-Fock component to 
dominate in Eq.(\ref{eq.pdfDefinition2}), the calculation of hadronic PDF's at moderade $x$ will require LFWFs which are the solution of infinite set of coupled integral equations, containing the different 
number of quark or gluon constituents.  Thus often in practical calculations LFWFs  are modeled. 
The modeling of the valence LFWFs for the pion has been done in different approaches such 
as Harmonic Oscillator models\cite{Dziembowski:1986dr}, models relating equal-time wave function 
to the light-front wave functions\cite{Huang:1994dy}, light-front consitutent quark models\cite{Pasquini:2014ppa}, the Nambu-Jona Lasinio model\cite{Dziembowski:1986dr} 
as well as model based  on light front holography \cite{Trawinski:2016jap,Trawinski:2016ygn}. 

Our approach in this work is phenomenological, instead of taking an infinite Fock state expansion as in Eq.(\ref{Fock-expansion}), we  approximate the $|h \rangle$ state through the transition of $\pi \rightarrow VR$, where $V$ is a valence $q\bar q$ core and $R$ is a residual state that absorbs all higher order Fock-states as well as possible configurations (such as null-mass components) that 
may couple with $q\bar q$-cluster. Furthermore, we define phenomenological wave functions 
in such a way that they represent the solution of the Weinberg type equation on the light-front for $q\bar q$ - cluster state (V), as well as  for cluster-residual (VR)  effective two body state.
Such wave functions are inherently non-perturbative and need to be modeled based on 
the prominent features of the structure of hadrons.

\section{The Pion in the Residual Field Model}
\label{pion-in-RMFM}

\subsection{General Formalism of Deep-Inelastic Scattering}
To set-up derivational framework for pion PDF we consider probing the pion through 
deep inelastic  electron scattering  ($e \pi \rightarrow e X $). Using the universality of PDFs 
the results of calculation can be compared with PDFs extracted for example from Drell-Yan processes in $\pi N\to l^+l^-+X$ reactions. This is a consequence of the definition of PDFs and validity of QCD factorization for these processes. 

Within one-photon exchange approximation the scattering takes place with the virtual photon $\gamma^*$, with four momentum $q^\mu = k_e^\mu - k_e^{\prime \mu}$, where $k_e^\mu$ and 
$k_e^{\prime \mu}$ are the four momenta of the initial and scattered electrons respectively.

The cross section for  the DIS process can be expressed through the contraction of 
leptonic and hadronic tensors as follows:
\begin{equation}
{d\sigma\over dQ^2 dx} = {2\pi\alpha^2 y^2\over Q^6}m_\pi L_{\mu\nu}W_\pi^{\mu\nu},
\end{equation}
where $Q^2 = -q^\mu q_\mu$, $y={p_\pi \cdot q\over p_\pi \cdot k_e}$ and the unpolarized leptonic tensor is 
\begin{equation}
L_{\mu\nu} = 4\left[\left(k_{e,\mu} - {q_\mu\over 2} \right)\left(k_{e,\nu} - {q_\nu\over 2} \right)\right]  + Q^2\left[ -g_{\mu\nu}- {q_\mu q_\nu\over Q2}\right].
\end{equation}
The hadronic tensor, $W_\pi^{\mu\nu}$,  in general can be defined through the inelastic hadronic  current $J^\mu(p_X, s_X,p_\pi,s_\pi)$ in the form:
\begin{eqnarray}
W_\pi^{\mu\nu} =  {1\over 4 \pi m_\pi} \int\sum\limits_X\sum\limits_{s_x} 
(J^\mu)^\dagger(p_X, s_X,p_\pi,s_\pi)J^{\nu}(p_X, s_X,p_\pi,s_\pi)
(2\pi)^4 \delta^4(q + p_\pi - p_X)\nonumber \\
\delta(p_X^2 - M_x^2) {d^4p_x\over (2\pi)^3}.
\label{wmunu0}
\end{eqnarray}

In derivations, we will work with light-front coordinates with the  convention for 4-momenta being:
\begin{equation}
k^\mu = (k^+, k^-, \mathbf{k}^\perp ) = \left(k^0 + k^3, k^0 -k^3, (k^1, k^2)\right).
\end{equation}

The calculations are performed  in the Drell-Yan-West reference frame where the 4-momenta of the pion, $p_\pi^\mu$ , and virtual photon, $q^\mu$ are defined as follows:
\begin{equation}
p_\pi^\mu = \left(p_\pi^+, \frac{m_\pi^2}{p_\pi^+}, \mathbf{0}^\perp \right), \ \ \ q^\mu = \left(0 , \frac{2 p_\pi\cdot q}{p_\pi^+}, \mathbf{q}^\perp\right),
\label{kins}
\end{equation}
where $q_\perp^2 = Q^2$.
 
In  this reference frame the $F_2$ structure function of the pion is related 
to  the $++$ component of the hadronic tensor as follows:
\begin{equation}
F_2(x,Q^2)  =  {m_\pi Q^2\over 2 x (p_\pi^+)^2}W_\pi^{+,+},
\label{F2Wmunu}
\end{equation}
and within the framework of QCD factorization and the definition of PDFs in  leading order  one can relate the PDF to the structure function $F_2$ as:
\begin{equation}
F_2(x,Q^2) = \sum\limits_i e_i^2 x f_i(x,Q^2),
\label{F2pdf}
\end{equation}
where $f_i(x)$ was defined in Eq.(\ref{eq.pdfDefinition2}) and $i$ runs for quarks or antiquarks in the pion, interacting with the external probe. Comparing 
Eqs.(\ref{F2Wmunu}) and (\ref{F2pdf}) one observes that the calculation of PDFs is 
related to the calculation of the  $++$ component of 
the hadronic tensor, $W_\pi^{\mu\nu}$.

In the case of valence quarks using the above relations allows us to 
extract sum of quark and anti-quark valence quark distributions weighted by $x$.
Usually, invoking isospin symmetry it is assumed that the up valence distribution in $\pi^+$ is equal to the down valence distribution in $\pi^-$: $f_V(x,Q^2) = u^{\pi^+}_V(x,Q^2) =d^{\pi^-}_V(x,Q^2)$.

\subsection{Hadronic Current and Hadronic Tensor}

\begin{figure}[htbp]
    \centering
         \includegraphics[width=0.95\linewidth, height=6cm]{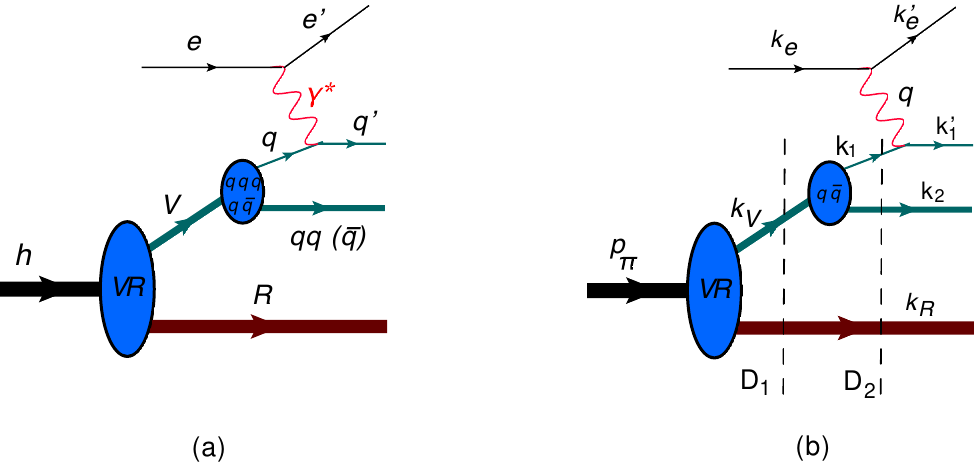}
     \caption{Feynman Diagrams for Deep Inelastic Scattering in Residual-Mean Field model.
    (a) General case for hadrons with $qqq$ or $q\bar q$ clusters. (b) For the case of deep inelastic scattering from pion.}
    \label{fig:rfm_diagrams}
\end{figure}

For calculation of pion structure function we consider the diagram of 
Fig.\ref{fig:rfm_diagrams}(b) in which it is assumed that the pion transitions to 
the cluster of valence $q\bar q$ state and the residual state, $R$  before 
the interaction of virtual photon with one of the valence quarks in the 
cluster takes place.

Identifying  the hadronic current $J^\mu$ in Eq.(\ref{wmunu0})  with the scattering amplitude $A^\mu$, which corresponds to  the part of the diagram of Fig.\ref{fig:rfm_diagrams}(b) without electron lines and virtual photon propagator and taking into account the phase space of final 
state quarks and residual system one obtains for the hadronic tensor: 
\begin{eqnarray}
W^{\mu\nu} (x, Q^2) = \frac{1}{4\pi m_\pi} \sum_{ \{h_i, \tau_i \}} 
\int  \delta(k_R^2- m_R^2)   \frac{d^4 k_R}{(2\pi)^3} 
\delta(k^{\prime,2}_1- m^2)   \frac{d^4 k^\prime_1}{(2\pi)^3}
\delta(k_2^2- m^2)           \frac{d^4 k_2}{(2\pi)^3}\nonumber \\
(2\pi)^4\delta^{(4)}(p_\pi+q - \sum_{i=1}^2 k_i -k_R)
 A^{\mu \dagger} A^\nu,
 \label{wmunu1}
 \end{eqnarray}
where the sum $ \sum_{ \{h_i, \tau_i \}} $ is over all the spins of the valence quarks and the residual system. The Lorentz invariant phase space evaluated  in light front coordinates is,
\begin{eqnarray}
\delta(k^2-m^2) d^4k & = &  \frac{1}{2} \delta(k^+k^- - k_\perp^2 -m^2) dk^- dk^+ d^2\mathbf{k}_\perp 
\nonumber  \\
& = &  \frac{ dk^+ d^2\mathbf{k}_\perp}{2k^+} |_{k^- = \frac{k_\perp^2 +m^2}{k^+}}
=  \frac{ dx d^2\mathbf{k}_\perp}{2x} |_{k^- = \frac{k_\perp^2 +m^2}{xP_\pi^+}} ,
\label{phasespace}
\end{eqnarray}
where $x =\frac{k^+}{P_\pi^+}$ is the light front momentum fraction of pion constituent with 
momentum $k^+$.
For energy-momentum conservation one obtains:
\begin{eqnarray}
 \delta^{(4)}(p_\pi+q -  k^\prime_1 - k_2 -k_R) = &&  2 
\delta (p_\pi^+ + q^+ - k^{\prime,+}_1 - k^+_2 - k_R^+) 
\delta (p_\pi^- + q^- - k^{\prime,-}_1 - k^-_{2}- - k_R^-)  \nonumber \\ 
&&\times \delta^{(2)}(\mathbf{p_\pi^\perp} + \mathbf{q}^\perp -  \mathbf{k^{\prime \perp}_1} - \mathbf{k_2^\perp} -  \mathbf{k}_{R}^{\perp}).
\label{delta4}
\end{eqnarray}
Taking into account our choice of the reference frame Eq.(\ref{kins}) according to which $q^+=0$ and $k^{\prime+}_1 = k_1^+$
one obtains:
\begin{equation}
\delta (p_\pi^+ + q^+ - k^{\prime+}_1 - k^+_2 - k_R^+) = \frac{1}{p_\pi^+} \delta (1 - x_1 - x_2-x_R),
\label{deltaplus}
\end{equation}
where $x_1 = {k_1^+\over p_\pi^+}$, $x_2 = {k_2^+\over p_\pi^+}$, and $x_R = {k_R^+\over p_\pi^+}$. 
Using $k_{1}' = p_\pi - k_2 - k_3 - k_R$ and neglecting initial transverse momenta of pion constituent, compared to $q_\perp$,i.e.  $k_{1, \perp}'^2 = (\mathbf{k}_{1, \perp} + \mathbf{q}_\perp )^2 \sim \mathbf{q}_\perp^2 \gg k_{1, \perp}^2$, as well as using the condition $p_\pi^+ \gg m_\pi, m_V,  \ m_R, \ k_{1,2}^\perp,  \ k_R$ one evaluates:

\begin{eqnarray}
\delta (p_\pi^- + q^- - k^{\prime-}_1 - k_2^- - k_R^-)  & = &  
\delta \left(\frac{m_\pi^2}{p_\pi^+}+ \frac{2 p_\pi \cdot q}{p_\pi^+} - \frac{k_{1,\perp}'^2 +m^2}{x_{1}p_\pi^+}  -  \frac{k_{2,\perp}^2 +m^2}{x_{2}p_\pi^+} -  \frac{k_{R,\perp}^2 +m_R^2}{x_{R}p_\pi^+} \right) \nonumber \\
&  \approx  & \delta \left(\frac{2 p_\pi \cdot q}{p_\pi^+} - \frac{Q^2}{x_{1}p_\pi^+}\right)  
 = \frac{x_{1}p_\pi^+}{2 p_\pi \cdot q} \delta(x_1 - x_B),
 \label{deltaminus}
\end{eqnarray}
where Bjorken $x$ is defined as $x_B = \frac{Q^2}{2 p_\pi \cdot q}$.  Finally with the same 
conditions:
\begin{equation}
 \delta^{(2)}(\mathbf{p_\pi^\perp} + \mathbf{q}^\perp -  \mathbf{k^{\prime \perp}_1} - \mathbf{k_2^\perp} -  \mathbf{k}_{R}^{\perp})=\delta^{(2)}(   \mathbf{k^{\perp}_1} +\mathbf{k_2^\perp} + \mathbf{k}_{R}^{\perp})
\label{deltaperp}
\end{equation}
Using Eqs.(\ref{deltaplus},\ref{deltaminus},\ref{deltaperp}) in Eq.(\ref{delta4}) for 
hadronic tensor in  Eq.(\ref{wmunu1}) one arrives at:
\begin{equation}
W^{\mu\nu} (x, Q^2) = \frac{1}{4m_\pi} \sum_{ \{h_i, \tau_i \}} \int  [dx][d^2 \mathbf{k}_\perp]  {2x_1x_B \over Q^2 }\delta(x_1 - x_B)
A^{\mu \dagger} A^\nu
\label{wmunu3}
\end{equation}
where  
\begin{eqnarray}
[dx] & = & \delta(1-  x_1 - x_2 - x_R) \frac{dx_R}{x_R}\prod_{i=1}^{2}  \frac{dx_i }{x_i}  \nonumber \\
\left[d^2 \mathbf{k}_\perp \right] & = &   
16 \pi^3 \delta^{(2)}\left(\sum_{i=1}^2 \mathbf{k}_{i,\perp} + \mathbf{k}_{R, \perp}\right)\frac{d^2 \mathbf{k}_{R, \perp}}{16 \pi^3 }\prod_{i=1}^{2}  \frac{d^2 \mathbf{k}_{i,\perp}}{16 \pi^3}.  
\end{eqnarray}
As it follows from Eq.(\ref{F2Wmunu}) and Eq.(\ref{F2pdf}) to calculate 
the pion PDF one needs to evaluate the "+" component of 
the scattering amplitude $A^\mu$.

\subsection{Scattering Amplitude in Residual Field Model}
To calculate the amplitude presented in Fig.(\ref{fig:rfm_diagrams})(b) we use light-front 
diagrammatic rules\cite{Lepage:1980fj}. Since momentum transferred to 
the quark/anti-quark in the pion is significantly larger than their initial momenta, the interference terms will be small and the cross section will represent the incoherent sum of the square of amplitudes corresponding to the scattering from quark and anti-quark. 
Thus we first calculate amplitude of scattering from quark and then use the 
result to evaluate the scattering from the anti-quark.

To calculate the amplitude we notice that the diagram of Fig.(\ref{fig:rfm_diagrams})(b) is characterized by two intermediate states (indicated by vertical dashed lines) that 
are defined by two  light-front propagators. With the momenta assigned for each 
particles in the scattering defined in the diagram one obtains:
\begin{equation}
A^\mu(k_1^\prime,h_1^\prime, k_2,h_2,k_R,h_R) = \sum\limits_{h_V} 
{\bar u(k_1^\prime, h_1^\prime) ie_q \gamma^\mu (\sh k_1 +m_q)
v(k_2,h_2)\Gamma^{V\to q\bar q}
\chi_V \over k_1^+ D_2}
{\chi_V^\dagger \chi_R^\dagger \Gamma^{\pi\to VR} \chi_\pi\over k_V^+ D_1},
\label{Amu}
\end{equation}
where $\Gamma^{\pi\rightarrow VR}$   is the 
effective vertex for the transition of pion into 
valence $q\bar q$ cluster, $V$ and a residual state $R$
and the $\Gamma^{V\rightarrow q\bar q }$ vertex describes the transition of the cluster, $V$ to 
the valence quark and anti-quark.
Also, $h_i$ and  $h_1^\prime$  are the helicities  of the i'th valence quark before and after the
interaction with virtual photon.

For the denominators $D_1$ and $D_2$ according to light-front diagrammatic rules, one has:
\begin{equation}
D_1 = p_\pi^- - k_R^- - k_V^- = {1\over p_\pi^+}\left(m_\pi^2 - {k_{R\perp}^2 + m_R^2\over x_R}- {k_{V\perp}^2 + m_V^2\over x_V}\right)
\label{D1}
\end{equation}
\begin{equation}
D_2 = p_V^-  - k_1^- - k_2^-- = {1\over k_V^+}\left(m_V^2 - {k_{1\perp}^2 + m^2\over \beta_1}- {k_{2\perp}^2 + m^2\over \beta_2}\right).
\label{D2}
\end{equation}
where  $\beta_i = \frac{x_i}{x_V}$ for $i=1,2$ quark and anti-quark.

Furthermore, we use LF sum rule for the numerator of the  propagator of off-shell quark with momentum $k_1$\cite{Lepage:1980fj}:
\begin{equation}
(\sh k_1 +m) = \sum\limits_{h_1}u(k_1,h_1)\bar u(k_1,h_1) + {\gamma^+ (k_1^2-m^2)\over k_1^+},
\label{propsum}
\end{equation}
and  notice that since we are interested in the $+$ component of the scattering 
amplitude, $A^\mu$, and  due to the relation $\gamma^+\gamma^+=0$, the second part of 
Eq.(\ref{propsum}) does not contribute to the $+$ component of the amplitude. 
Inserting the first part 
of Eq.(\ref{propsum}) into Eq.(\ref{Amu}) one introduces a phenomenological 
LF wave function for the Valence-Residual system:
\begin{equation}
    \psi_{VR}(x_V,\mathbf{k}_{R,\perp},x_R,\mathbf{k}_{V,\perp}) = \frac{\bar\chi_V\bar \chi_R\Gamma^{\pi\rightarrow VR}}{m^2_{\pi} -\frac{k^2_{V,\perp}+m^2_V}{x_V}-\frac{k^2_{R,\perp}+m^2_R}{x_R}}
\end{equation}
and similarly for the $q\bar q$ cluster:
\begin{equation}
    \psi_{q\bar q}(\beta_1,\beta_2,\mathbf{k}_{1,\perp},\mathbf{k}_{2,\perp}) = \frac{\bar u(k_1,h_1)\Gamma^{V\rightarrow q\bar q }v(k_2,h_2)\chi_V}{m^2_V -\sum_{i =1}^2\frac{k^2_{i,\perp}+m^2_i}{\beta_i}},
\end{equation}
where $\chi_V$ and $\chi_R$ represent the helicity wave functions for the cluster and residual 
system respectively.

In further derivation we assume that the residual state has a zero helicity and the quantum number of $q\bar q$-cluster  
is defined by  the quantum number of the pion.  Writing then the helicity wave function of the  cluster explicitly:  $\xi_{q\bar q} = {u(k_1,h_1){\bar v}(k_2,-h_1) -  
u(k_1,-h_1)\bar v(k_2,h_1)\over \sqrt{2}}$, 
for the "+"-component of the scattering amplitude one obtains:
\begin{eqnarray}
&&A^+(k_1^\prime,h_1^\prime, k_2,-h_1,k_R) = \left[\bar u(k_1^\prime, h_1^\prime) ie_q \gamma^+ u(k_1,h_1)  {\psi_{q\bar q}(\beta_1,\beta_2,\mathbf{k}_{1,\perp},\mathbf{k}_{2,\perp})\over \sqrt{2}\beta_1}\right.
\nonumber \\
&&
\left.
-   \bar v(k_1,h_1)ie_{\bar q} \gamma^+ v(k_1^\prime, h_1^\prime)  {\psi_{q\bar q}(\beta_1,\beta_2,\mathbf{k}_{1,\perp},\mathbf{k}_{2,\perp})\over \sqrt{2}\beta_1}\right]
 {\psi_{VR}(x_V,\mathbf{k}_{R,\perp},x_R,\mathbf{k}_{V,\perp})\over x_V}.
\end{eqnarray}
By neglecting the masses of valence quarks, one evaluates only helicity conserving 
term of the matrix elements of: $\bar u(k_1^\prime, h_1^\prime)\gamma^+ u(k_1,h_1)  
=  \bar v(k_1,h_1) \gamma^+ v(k_1^\prime, h_1^\prime) = 2k_1^+\delta^{h^\prime_{1},h_1}$.
This results in:
\begin{eqnarray}
 A^+(k_1^\prime,h_1, k_2,-h_1,k_R) = &&   \left[2k_1^+ ie_q
{\psi_{q\bar q}(\beta_1,\beta_2,\mathbf{k}_{1,\perp},\mathbf{k}_{2,\perp})\over \sqrt{2}\beta_1} 
-  2k_1^+ ie_{\bar q}    {\psi_{q\bar q}(\beta_1,\beta_2,\mathbf{k}_{1,\perp},\mathbf{k}_{2,\perp})\over \sqrt{2}\beta_1}\right]
\nonumber \\
&&\times {\psi_{VR}(x_V,\mathbf{k}_{R,\perp},x_R,\mathbf{k}_{V,\perp})\over x_V}.
\label{A+}
\end{eqnarray}

\subsection{Valence Parton Distribution}
Inserting Eq.(\ref{A+}) into the expression of the hadronic tensor in Eq.(\ref{wmunu3}), summing by helicites of final quarks and 
using relations of Eq.(\ref{F2Wmunu}) and (\ref{F2pdf}) for the valence quark distribution one obtains:

\begin{eqnarray}
 f_V(x_B, Q^2) & = &  \int \frac{dx_1 d^2 \mathbf{k}_{1, \perp}}{16\pi^3 x_1} \frac{dx_2 d^2 \mathbf{k}_{2, \perp}}{16\pi^3x_2}  \frac{dx_Rd^2 \mathbf{k}_{R, \perp}}{16\pi^3 x_R}\delta(x_1-x_B)   16 \pi^3 \delta \left(1- x_1-x_2 - x_R\right)
\nonumber \\
& \times &  \delta^{(2)}\left(\sum_{i=1,2,R}\mathbf{k}_{i, \perp}\right)   
\mid  \Psi_{q\bar q}(\beta_1,k_{1,\perp},\beta_2,k_{2,\perp})\mid^2 
\mid \Psi_{VR}(x_R,k_{R,\perp})\mid ^2.
\label{pdf1}
\end{eqnarray}
Expecting that Light-Front wavefunction of the $q\bar q$ system will depend on 
relative transverse momentum and relative light-front momentum fractions only, 
we can factorize above integrals by introducing transverse momentum of quark and anti-quark 
in the center of mass of the $q\bar q$ system
 \begin{equation}
\tilde{\mathbf{k}}_{i, \perp} = \mathbf{k}_{i, \perp} - \beta_i \mathbf{k}_{V, \perp}  \ \ \ \ \text{(i=1,2)}.
\label{krel}
\end{equation}
Substituting Eq.(\ref{krel}) into Eq.(\ref{pdf1}) one can factorize the integration by 
$k_{R,\perp}$. Furthermore integrating over $x_2$ results in 
\begin{eqnarray}
& & f_V(x_B, Q^2)   =     
\int_0^{1-x_B} \frac{dx_R }{(16\pi^3)^2 x_B (1 - x_B - x_R )x_R}\nonumber \\
& & \ \ \ \ \ \  \times \left[  
\int^{Q^2} d^2\tilde{\mathbf{k}}_{1, \perp}d^2\tilde{\mathbf{k}}_{2, \perp}
\delta^2 (\tilde{\mathbf{k}}_{1, \perp} +\tilde{\mathbf{k}}_{2, \perp}) 
|\Psi_{q \bar q}(\beta_1, \tilde{\mathbf{k}}_{1, \perp})|^2
\int^{Q^2} d^2 \mathbf{k}_{R, \perp}|\psi_{VR}(x_R, \mathbf{k}_{R, \perp})|^2\right].\nonumber \\
\label{pdf3}
\end{eqnarray}
The above expression can be integrated 
by $d^2\tilde k_{2,\perp}$ using the $\delta^2()$ function and 
renaming $\beta_1 \equiv \beta$ and $ k_{1,\perp}\equiv k_\perp$ one obtains:
\begin{eqnarray}
& & f_V(x_B, Q^2)   =     
\int_0^{1-x_B} \frac{dx_R }{(16\pi^3)^2 x_B (1 - x_B - x_R )x_R}\nonumber \\
& & \ \ \ \ \ \  \times \left[  
\int^{Q^2} d^2\mathbf{k}_{\perp} |\Psi_{q \bar q}(\beta, {\mathbf{k}}_{\perp})|^2
\right ]_{q\bar q}
\left[\int^{Q^2} d^2 \mathbf{k}_{R, \perp}|\psi_{VR}(x_R, \mathbf{k}_{R, \perp})|^2\right]_{VR},\nonumber \\
\label{pdf4}
\end{eqnarray}
where $[\cdots]_{q\bar q}$ and $[\cdots]_{V R}$ we can interpret as 
response functions of the $q\bar q$ valence-cluster  and residual systems, respectively.

\section{Modeling Nonperturbative Light-Front Wave Functions}\label{Sec:Model_LFWF}

\subsection{General approach for two-body relativistic LFWF}

To model $q\bar q$ cluster and $VR$ wave functions we first note that both 
are two-body wave functions for which relativistic normalization is defined 
as follows:
\begin{equation}
\int \mid \Psi^{LF}(k_{12})\mid ^2 {dy\over y(1-y)}{d^2k_{12,\perp}\over 16\pi^3} = N,
\label{relnorm}
\end{equation}
where 
\begin{equation}
y = {k_1^+\over p_T^+},
\end{equation}
with $k_1^\mu$ and $p_T^\mu$ being four-momenta of one of the bound particles and 
the target.
Here the relative momentum in the CM frame of two particle system is:
\begin{eqnarray}
&&k_{12}^2 = \frac{\left(s_{12} -(m_1-m_2)^2\right)\left(s_{12}-(m_1+m_2)^2\right)}{4s_{12}}    \nonumber \\
&&= {m_1^2(1-y)+ m_2^2y + k^2_{12,\perp}\over 4 y(1-y)} - {m_1^2+m_2^2\over 2}
+ {(m_1^2 - m_2^2)^2y(1-y)\over 4(m_1^2(1-y) + m_2^2 y + k^2_{12\perp})},
\label{relmom}
\end{eqnarray}
and
\begin{equation}
s_{12} = {m_1^2(1-y) + m_2^2y + k^2_{12,\perp}\over y (1-y)}.
\end{equation}

We model LF wave function similar to \cite{Frankfurt:1981mk,Sargsian:2022rmq} 
assuming factorization of 
the relative momentum  wave function in the CM system and 
the absence of inelastic transitions in two-body wave function. This allows to relate 
the CM wave function to the non-relativistic 
wave function evaluated at the relative momentum defined relativistically in Eq.(\ref{relmom}):
\begin{equation}
\Psi_{12}^{LF}(k_{12}) = \Psi_{12}^{NR}(k_{12}) 
\sqrt{ {E_1E_2\over E_1+E_2}16\pi^3}.
\label{LFWFmodel}
\end{equation}
Here non-relativistic wave function is normalized as follows:
\begin{equation}
\int \mid \Psi^{NR}(k_{12})\mid ^2 d^3k_{12} = N.
\label{NRnorm}
\end{equation}
It is worth noting that due to relativistic normalization in Eq.(\ref{relnorm}) and 
definition of the light-front relative momentum $k_{12}$ in Eq.(\ref{relmom}, the 
$k_{12}\to \infty$ corresponds to the limits of  $y\to 0, 1$, 
thus excluding non-physical region of momenta entering in the argument of LF wave function
(as it is the case of non-relativistic wave functions).

The normalization factor $N$, in Eqs.(\ref{relnorm}) and (\ref{NRnorm}) in general should 
be fixed through observables, such as baryonic number (or number of constituents)\cite{Frankfurt:1985ui,Jaffe:1988up}, or 
form-factor of the system\cite{Sargsian:2009hf}, etc. For example for the case of baryons the valence quark wave 
function's normalization can be fixed by the baryonic number\cite{Leon:2020bvt}. However, 
a situation with pion introduces some uncertainty since the baryonic number is zero and this issue will be discussed below.

\subsubsection{Soft $q \bar q$ Wave Function}
Assuming quark masses are the same, the above approach can be used to model the soft $q\bar q$ wave function in the 
form:
\begin{equation}
\Psi^{LF}_{q\bar q}(\beta,k_{12,\perp}) = \sqrt{8\pi^3 E_k}\Psi^{NR}_{q\bar q}(k_{12}),
\label{wfqqbar}
\end{equation}
where, $E_k = \sqrt{m^2 + k_{12}^2}$  and 
$\vec k_{12} = (k_{12,z},\vec k_{12,\perp})$ is the relative momentum 
in the $q\bar q$ CM system, with the magnitude
\begin{equation}
k_{12} = \sqrt{{m^2 +  k^2_{12,\perp}\over 4 \beta(1-\beta)} - m^2}.
\label{k12}
\end{equation}
We normalize the $q\bar q$ wave function  per one quark (or anitquark) in the cluster since in the calculation of the 
PDF we sum by individual contributions of valence quarks.
We  also take  
into account that the normalization is less than unity since no hard component is included in the cluster wave function 
which can be generated through the hard gluon exchanges\cite{Lepage:1980fj}. With this, the normalizetion is fixed as:
\begin{equation}
\int \mid \Psi^{LF}{q\bar q}(k_{12})\mid ^2 {d\beta\over \beta(1-\beta)}{d^2k_{12,\perp}\over 16\pi^3} = N_{q\bar q} =1-h.
\label{qbarqnorm}
\end{equation}
Note that in addition to the hard component the parameter $h$ also accounts for the fact that  at $x<0.1$, the Regge dynamics 
may play a role in PDF which is not included in current calculations.

For non-relativistic wave function we use a 
Harmonic Oscillator type  wave function with free parameters, $A_V$ and $B_V$:
\begin{equation}
\Psi^{NR}_{q\bar q}(k) = A_V \exp{\left[-\frac{B_V}{2} k_{12}^2\right]}
\label{qbarqNR}
\end{equation}

A number of previous works have used the similar wave functions or a form similar to it \cite{Brodsky:1981jv,Lepage:1980fj, Gunion:1983ay, Trawinski:2016jap, Pasquini:2018oyz}.
Using the above light-front wave function, Eq.(\ref{wfqqbar}), for the $q\bar q$ system's
response function defined in Eq.(\ref{pdf4}), one obtains:
\begin{eqnarray}
&& \left[  \int^{Q^2} d^2\mathbf{k}_{\perp} |\Psi_{q \bar q}(\beta, {\mathbf{k}}_{\perp})|^2
\right ]_{q\bar q} 
= 8\pi^3\int^{Q^2}d^2k_\perp E_k A_V^2 e^{-B_V{k_\perp^2\over 4\beta(1-\beta)}}\nonumber \\
&& = 8\pi^3\int^{Q^2} d^2\mathbf{k}_{\perp}\sqrt{{k_\perp^2\over 4\beta(1-\beta) }}
e^{-B_V{k_\perp^2\over 4\beta(1-\beta)}} = {16\pi^4 \sqrt{\pi} A_V^2\over B_V^{3\over 2}}\beta(1-\beta) = 
16\pi^3 N_{q\bar q} \beta(1-\beta).\nonumber \\
\label{responseqbarq}
\end{eqnarray}
Here in the last step we neglected  by the masses of the valence quarks and assumed 
$Q^2\gg k_\perp^2$ (consistent with collinear approximation). This allowed to calculate the integral analytically, in which  we set the upper  limit of the integration to $\infty$.  It is interesting that in the massless limit because of the relation of Eq.(\ref{k12}) 
the integral in the r.h.s part of 
the equation has an analytic form  of the normalization of the NR wave function in the center of mass of $q\bar q$ system. 
As a result the response function depends on the overall normalization factor $N_{q\bar q} = {\pi^{3\over 2}A_{V}^2 \over B_V^{3\over 2}}$ while momentum distribution, 
$\beta(1-\beta)$ is similar to the one obtained from perturbative wave function.

It is worth mentioning that the calculated in this way  $q\bar q$'s response function peaks at $\beta={1\over 2}$ which is consistent with relativistic wave function of massless $q\bar q$ system.

\subsubsection{Light-Front Wave Function of Valence-Residual System} 
To calculate the response function of residual system one has to model 
the light-front wave function of valence-residual system.
Here we use a Harmonic Oscillator (HO) type wave function in Eq.(\ref{LFWFmodel}) obtaining
\begin{equation}
\Psi_{R}^{LF}(x_R, k_{R,\perp}) = 
A_{R}\sqrt{16\pi^3}\sqrt{ {E_VE_R\over E_V+E_R}}e^{-{B_{R}\over 2} k_{R}^2}, 
\label{LFVR}
\end{equation}
where $E_V = \sqrt{m_V^2 + k_{R}^2}$, $E_R = \sqrt{m_R^2 + k_{R}^2}$. The $k_R$ is the 3-dimensional momentum defined in the center of mass of the $V-R$ system:
\begin{equation}
k^2_{R} = \frac{(s_{VR}-(m_V - m_R)^2)(s_{VR}-(m_V + m_R)^2)}{4s_{VR}},
\label{kR}
 \end{equation}
where 
\begin{equation}
s_{VR} =  \frac{k_{R,\perp}^2 +m_R^2}{x_R} +  \frac{k_{R,\perp}^2 +m_V^2}{1-x_R},
\label{SVR}
\end{equation}
resulting in 
\begin{eqnarray}
k_{R}^2 = &&
{m_R^2(1-x_R)+ m_V^2x_R + k^2_{R,\perp}\over 4 x_R(1-x_R)} \nonumber \\ &&-{m_R^2+m_V^2\over 2}
+ {(m_R^2 - m_V^2)^2x_R(1-x_R)\over 4(m_R^2(1-x_R) + m_V^2 x_R + k^2_{R\perp})}.
\end{eqnarray}
The above defined wave function is normalized as follows:
\begin{equation}
\int \mid \Psi^{LF}_{R}(k_{12})\mid ^2 {dy\over x_R(1-x_R)}{d^2k_{R,\perp}\over 16\pi^3} = N_R,
\label{Rnorm}
\end{equation}
where the normalization, $N_R$ one can interpret as  the number of constituents in the residual hadronic system.

Using the above expressions for the $VR$ wave function and the relative momentum $k_R$~(Eq.(\ref{kR})), for the response function of the $VR$ system, one obtains the following:
\begin{equation}
R(x_R)= \left[\int^{Q^2} d^2 \mathbf{k}_{R, \perp}|\psi_{VR}(x_R, \mathbf{k}_{R, \perp})|^2\right]_{VR} = {16\pi^3}A_R^2 \int^{Q^2} d^2 \mathbf{k}_{R, \perp} 
{E_VE_R\over (E_V+E_R)}e^{-B_R k_{R}^2}.  
\label{responseVR}
\end{equation}
Note that in the above equation  masses of the valence cluster $m_V$, 
and residual state $m_R$, are virtual and characterize the  transition into 
non-stationary virtual state of a valence cluster and the residual system, thus 
they do not need to correspond to the pion mass.   

\subsection{Modeled Soft  Pion PDF}
Inserting Eq.(\ref{responseqbarq}) and Eq.(\ref{responseVR}) into Eq.(\ref{pdf4})  
and using normalization codition of Eq.(\ref{qbarqnorm}, for 
the modeled parton distribution of pion one obtains:
\begin{eqnarray}
& & f_V(x_B, Q^2)   =   
{N_{q\bar q}A_{R}^2}
 \int_0^{1-x_B} \frac{dx_R }{x_R} {1\over (1-x_R)^2}\ 
 \left[\int^{Q^2} d^2 \mathbf{k}_{R, \perp} 
{E_VE_R\over (E_V+E_R)}e^{-B_R k_{R}^2}  \right].  \nonumber \\
\label{pdf3}
\end{eqnarray}
The obtained expression indicates that,  the $\beta(1-\beta)$- relative light-cone momentum fraction  
distribution of valence quarks in the cluster 
(Eq.(\ref{responseqbarq}))  results in 
an only  $x_R$ dependent  integrand in  Eq.(\ref{pdf3}) convoluted with the 
response function of the residual system (Eq(\ref{responseVR})).
 In the current model this follows from the neglecting the valence quark masses.
As Eq(\ref{pdf3}) shows the shape of the PDF depends on the slope factor of the 
cluster-residual system's momentum distribution, $B_R$ as well as the respective virtual masses of the cluster $m_V$ and the residual system $m_R$.

\section{Numerical Results}\label{fitting-results}
\subsection{Parameters of the model}

We use recent results from JAM collaboration\cite{barry2018first,barry2021global}  on 
pion PDF to fit Eq.(\ref{pdf3}).  Our main focus in the fitting procedure is to
describe the position and the height of the x weighted parton distribution in accordance with our model assumption that these features of valence PDFs are characterized by the soft partonic dynamics. As follows from Eq.(\ref{pdf3}) the masses of residual, $m_R$, 
and valence, $m_V$,  systems, as well as the width of the $VR$ distribution, $B_R$ define  the  shape of the PDF that produces  the   peaking distribution of 
the PDF weighted by $x$.  

We choose the starting $Q^2= Q_0^2 = 1.69$~GeV$^2$, where $Q_0$ is chosen in JAM parameterization to be the $c$-quark mass.
The fitting analysis indicates that the position of the peak can be reproduced 
only if the virtual mass of the residual system is close to zero, $m_R\approx 0$.  
This is very different from our analysis of nucleon PDF's in residual field model\cite{Leon:2020cev,Leon:2020nfb} in which we found that the residual mass for the best fit is close to the pion mass, $m_R^N\approx 0.14$, indicating the validity of meson cloud picture of valence parton distribution in the nucleon.

\begin{figure}
     \centering
     \includegraphics[width=0.8\linewidth]{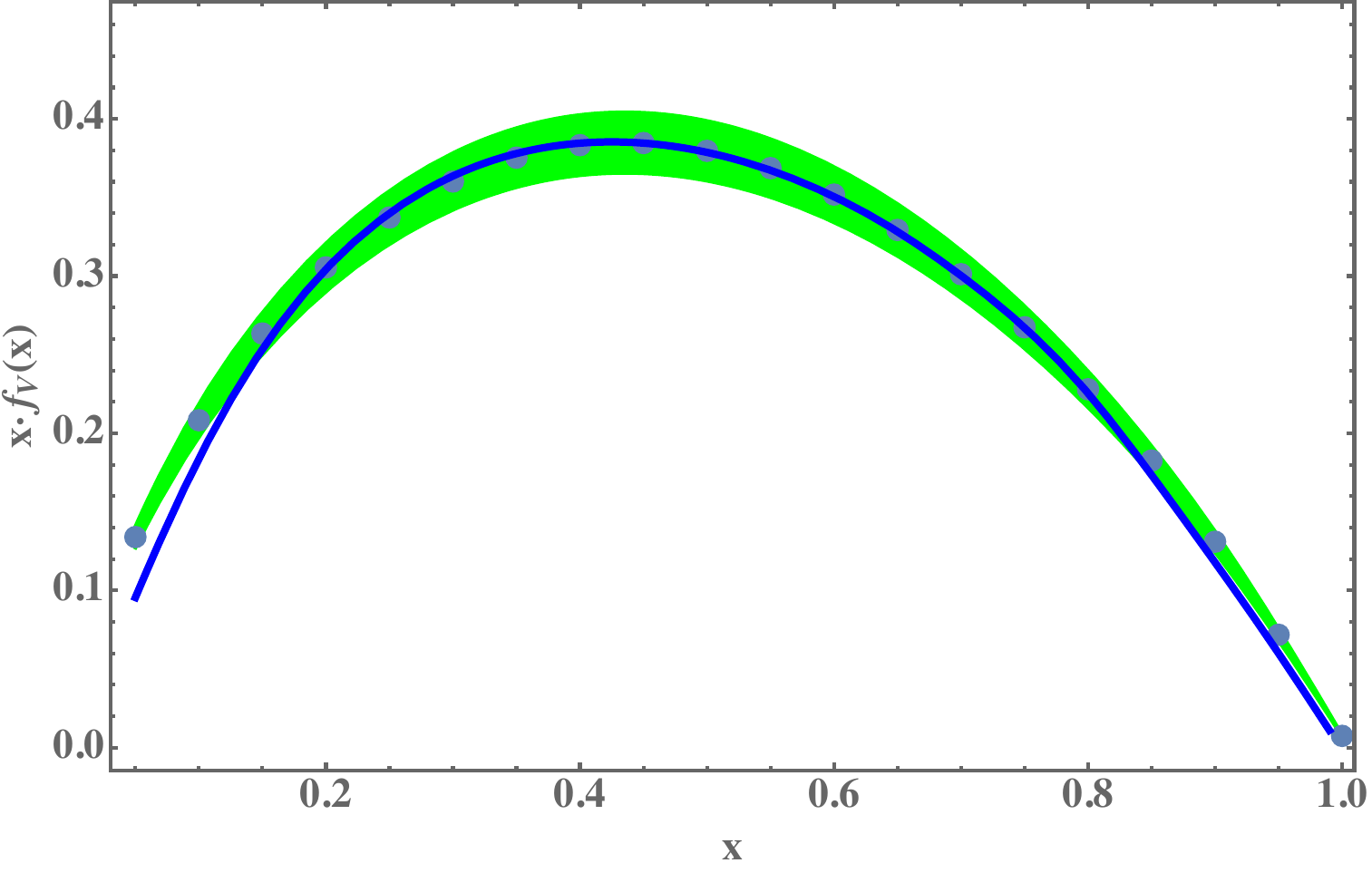}
     \caption{Description of x-weighted pion PDF at $Q_0=1.3$~GeV for $m_R=0$, $m_V=0.95$~GeV and $B_R = 1.2$~GeV$^{-2}$.}
     \label{fig:pionFit}
 \end{figure}

The best fit of the JAM pion PDF is presented in Fig.\ref{fig:pionFit}, for the 
parameters of $m_R=0$, $m_V= 0.95$~GeV and $B_V = 1.2$GeV$^{-2}$.  Note that the qualitative agreement of the data requires $m_V\sim Q_0$ and $m_R\sim 0$. As the figure shows, the soft mechanism describes the empirical PDF reasonably well in the range of $0.1< x < 0.85$, leaving very little room for the hard mechanism in the  generation of PDF at large x.

 \subsection{Interpretation of the Results}
\begin{figure}
    \centering
    \includegraphics[width=0.8\linewidth]{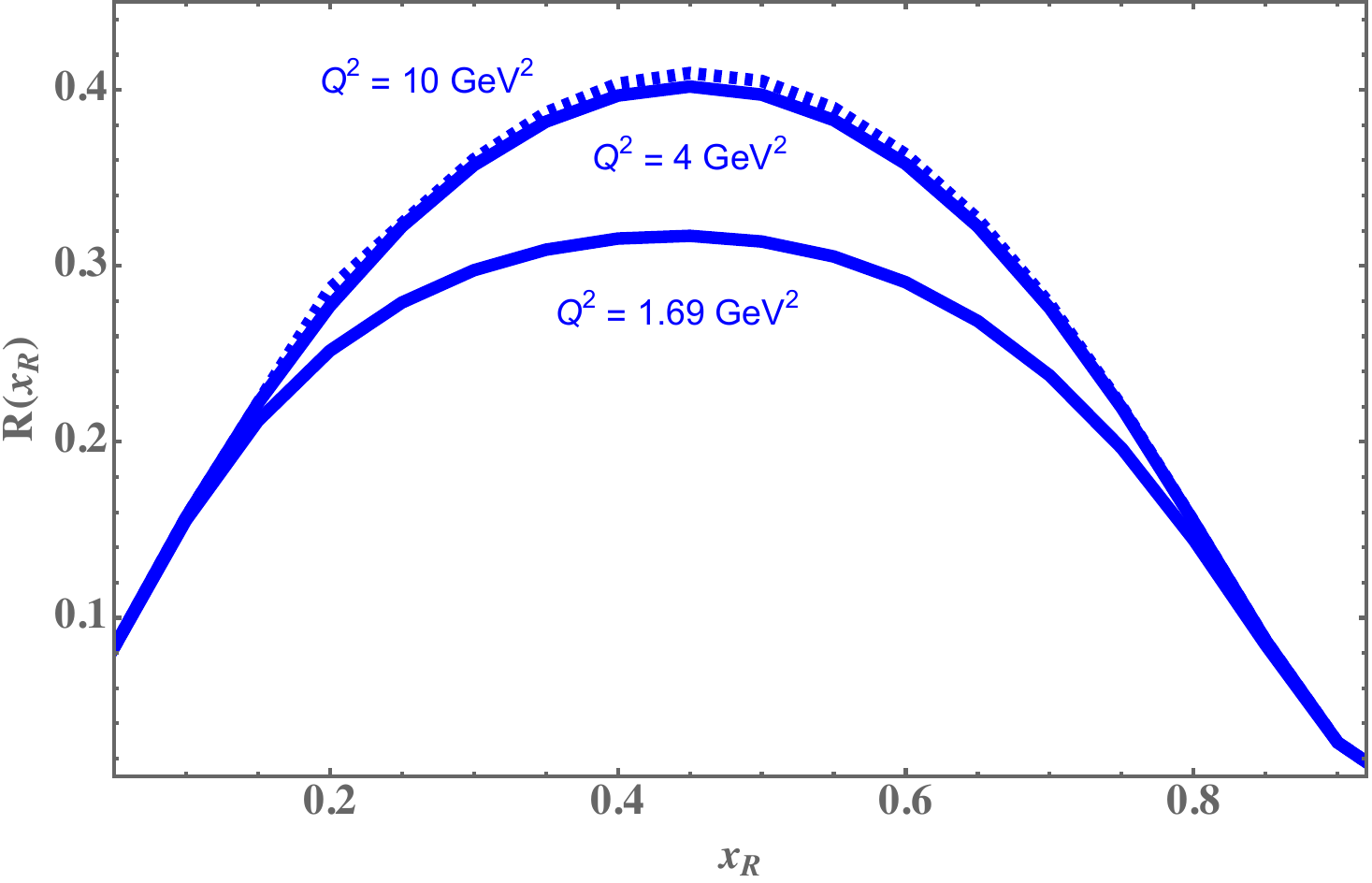}
    \caption{The $x_R$ dependence of response function $R(x_R)$ at different $Q^2$.      }
    \label{fig:ResponseVR}
\end{figure}

 The parameters of the model that provide the best fit to the empirical PDF allow to 
 ascertain the mechanism dominating parton dynamics at the x-range  where the peak of 
 x-weighted PDF distribution is observed. For this, we first analyze the response function 
 of the residual system presented in Eq.(\ref{responseVR}). As Fig.\ref{fig:ResponseVR} shows  the response function at starting $Q_0$, has a wide distribution in $x_R$, slightly peaking at $x_R\approx 0.5$.  The distribution saturates at higher $Q^2$ with 
 the peaking at $x_R = 0.5$  becoming more pronounced. This indicates that the valence $q\bar q$ system as a whole carries a total momentum fraction $x_V = 1-x_R\approx 0.5$.

Since $x$-Bjorken corresponds to the light-front momentum fraction of 
the interacting quark with the relation $x\approx \beta(1-x_R)$, where $\beta$ is 
the LF momentum fraction of the $q\bar q$ cluster carried by the interacting quark, and 
since the broad distribution of $x_R$,
one observes that  for $x\gtrsim 0.5$, relative momentum fraction, $\beta\sim 1$.  
This indicates that  the observed distribution  favors the dominance of the Feynman 
mechanism\cite{Feynman:1973xc} according to which the interacting quark carries the most 
of the light-front momentum fraction  of the cluster, with remaining LF momenta 
distributed among the  wee partons in the residual system.

Another observation (as  noted above)  is the result of $m_R\approx 0$. 
This indicates that  the residual system does not exhibit any structure 
(as it was the case for valence PDFs of nucleons \cite{Leon:2020bvt}). 
On the other hand, the large value of  $m_V\approx 0.95$~GeV indicates the large virtuality of  the 
interacting quark comparable to $Q_0$ of the probe. This together with the  relatively 
large $B_R$ indicates that while interacting valence quark is highly virtual, 
the residual system is spread over the bulk size of the pion.
This picture is consistent with the phenomenology that nucleon-nucleon interactions at large distances ($\ge 1.5fm$) 
proceeds through  the pion exchange in which no quark-interchanges are involved. 

The overall  constant in Eq.(\ref{pdf3}), is evaluated at  
$\sqrt{N_{q\bar q}} A_R = \sqrt{1-h}A_R \approx 0.86$. Assuming that overall Regge and high momentum 
mechanism contribute to $h\approx 0.2$, one obtains for the normalization of the residual LF wave function 
in Eq.(\ref{Rnorm}) $N_R\approx 4$.  This will predict, for example, that integrated multiplicity for 
pion in the target fragmention region to be above $2$.

Finally, the surprising result of the fit is that the considered model leaves little room for 
hard mechanism of high x PDF generated through the hard gluon exchange between valence $q\bar q$ 
system. The current result favors a pion structure dominated by soft mechanism of valence quark interaction.
Moreover, it is interesting that $f_V(x)\mid_{x\to 1} \sim (1-x)$ consistent with the 
$f_V(x) \sim 1-x$ distribution at the intermediate region in which case $xf_V(x) \sim x(1-x)$ peaks at $x_p \sim 0.5$ in agreement  with the data. The later reinforces that observation that high $x$ structure of pion PDF has large soft 
component that defines also the peaking properties of $xf_V(x)$ distribution.

\section{Conclusion and Outlook}\label{conclusion-and-outlook}
We have calculated pion valence parton distribution within  residual field model, which was
previously applied to the calculation of nucleon PDFs. The model assumes no hard 
interaction between valence quarks and it separates $q\bar q$ valence cluster from 
the residual system.  Both, $q\bar q$ and $VR$ systems are described by nonperturbative 
light-front wave functions that satisfies relativistic normalization. These wavefunctions 
allow  to introduce the residual response function as well as the response function of the $q\bar q$ valence cluster. 
The pion PDF represents the convolution of these response functions. 
We analyzed the parameters of the model by fitting the position and the height of the 
x-weighted PDF using  empirical JAM18 distribution. 

The fitting indicates that one can describe the above attributes of pion PDF only if 
the mass of the residual system is close to zero. However, we obtain large virtual mass for 
the $q\bar q$ cluster comparable to the $Q_0$ of the probe. Analysis of the response function 
of the residual system indicates that the dominant mechanism is similar to the Feynman mechanism in which interacting parton carries almost all the momentum of the valence quark system, with the remaining light-front momentum distributed to the non-interacting wee partons.

The fit allows to describe the empirical PDF reasonably well in the range of $0.1\le x \le 0.85$
leaving little room for the hard mechanism of generating PDFs at large x. 
The model practically resolves the observed $(1-x)^\beta$, $\beta\approx 1$ scaling relation of 
the pion PDFs at large x.  The large soft part contribution at high x, indicates that the introduction of the hard gluon exchange mechanism in  the generation of 
high x PDF will require the consideration of the interference between soft and hard amplitudes that can introduce 
the exponent, $\beta<2$ in $(1-x)^\beta$ distribution at very large x.

With the parameters of light-Front wave function of residual system fixed, 
one will be able to make numerical predictions for different observables such as electromagnetic form-factors,  generalized parton distribution functions (GPDs) or transverse momentum dependent parton distribution functions (TMDs) in pion that will allow an additional  verification of the model. Additionally, the modeled residual state wave function allows to 
evaluate the fragmentation of pion in the backward region that can be probed in the EIC collider kinematics.

\section*{Acknowledgement}
Authors are thankful to Alan Sosa for useful discussions.
This work is supported by United States Department of Energy grant under contract DE-FG02-01ER41172.  

%

\end{document}